# Understanding the genetic basis of variation in meiotic recombination: past, present, and future.

Susan E. Johnston.
Institute of Ecology and Evolution, School of Biological Sciences, University of Edinburgh, Edinburgh EH9 3FL, United Kingdom.
Susan.Johnston@ed.ac.uk

## Abstract

Meiotic recombination is a fundamental feature of sexually reproducing species. It is often required for proper chromosome segregation and plays important role in adaptation and the maintenance of genetic diversity. The molecular mechanisms of recombination are remarkably conserved across eukaryotes, yet meiotic genes and proteins show substantial variation in their sequence and function, even between closely related species. Furthermore, the rate and distribution of recombination shows a huge diversity within and between chromosomes, individuals, sexes, populations, and species. This variation has implications for many molecular and evolutionary processes, yet how and why this diversity has evolved is not well understood. A key step in understanding trait evolution is to determine its genetic basis - that is, the number, effect sizes, and distribution of loci underpinning variation. In this perspective, I discuss past and current knowledge on the genetic basis of variation in recombination rate and distribution, explore its evolutionary implications, and present open questions for future research.

## Introduction

In sexually-reproducing species, homologous recombination is required for proper segregation of chromosomes during gametogenesis (Koehler et al. 1996). It also shuffles alleles on the same chromosome into novel combinations, making it an important evolutionary force for both adaptation and maintenance of genetic diversity in populations (Felsenstein 1974; Otto and Lenormand 2002). Recombination is the result of DNA double strand break (DSB) repair during meiosis that can be resolved in two ways: crossovers, where large sections are reciprocally exchanged between homologous chromosomes; and gene conversions, where a given DSB in one homologue is repaired by its sister homologue, resulting in the non-reciprocal transfer of short tracts of DNA (reviewed in Wang et al. 2015; Lorenz and Mpaulo 2022). The mechanisms of recombination are remarkably conserved into the deep evolutionary past of animals, plants, and fungi, with homologous proteins involved in the processes of chromosome pairing, assembly of the synaptonemal complex (a protein structure that forms between paired chromosomes), DSB formation and repair via homologous recombination, and designation of crossover and non-crossover sites (reviewed in Arter and Keeney 2023). Despite this, the genes and proteins involved in these processes show remarkable variation in their sequence and function even between closely related species (Gerton and Hawley 2005; Kumar et al. 2010; Keeney et al. 2014; Dapper and Payseur 2019; Arter and Keeney 2023).

This is reflected in the vast diversity in the rate and distribution of recombination events across eukaryotes, which is manifest in different ways. The rate and distribution of recombination often vary within and between taxonomic groups, species, and populations (Price and Bantock 1975; Stapley et al. 2017; Samuk et al. 2020). They can vary between individuals of the same population, often due to genetic differences between them (Kong et al. 2014; Brekke et al. 2023). They can be markedly different between females and males in the same species (Lenormand and Dutheil 2005), and even in female and male gametes from the same hermaphroditic individual (Theodosiou et al. 2016). They also vary relative to broad and fine-scale features of the genome (Haenel et al. 2018; Brazier and Glémin 2022), and often occur in narrow recombination "hotspots" throughout the genome (reviewed in Choi and Henderson 2015; Zelkowski et al. 2019). Finally, they can be plastic, varying with individual and environmental factors such as age, temperature, infection status, and oxidative stress (reviewed in Bomblies et al. 2015; Rybnikov et al. 2023).

Understanding the causes and consequences of this diversity has been of interest to molecular and evolutionary biologists for more than a century (Box 1). This is because recombination occurs at a critical point in the life cycle, and its direct outcomes may impact individual fertility and the health of offspring (Hassold and Hunt 2001; Berg et al. 2010; Lachance and Tishkoff 2014). Furthermore, many evolutionary processes are influenced by recombination, including aspects of adaptive evolution (Gossmann et al. 2014; Castellano et al. 2016; Kosheleva and Desai 2018; Rousselle et al. 2019), effects of background selection (Booker et al. 2022), mutation rates (Duret and Arndt 2008; Hinch et al. 2023), sex chromosome evolution (Charlesworth 2017; Olito and Abbott 2023), hybridisation and the fate of introgressed alleles (Martin and Jiggins 2017; Schumer et al. 2018; Duranton and Pool 2022), and speciation (Ortiz-Barrientos et al. 2016). However, understanding how and why variation in recombination *itself* has evolved remains a constant challenge, and there are still open questions on if recombination is adaptive, stochastic, and/or evolving as a consequence of selection on correlated traits (reviewed in Lenormand et al. 2016; Ritz et al. 2017; Henderson and Bomblies 2021). Whilst much progress has been made, quantifying recombination has been challenging, and much of our understanding was limited to a handful of model species (Box 1). Nevertheless, advances in technologies and methodologies over the last decades have led to new insights accumulating at an impressive pace across a diverse range of systems (reviewed in Stapley et al. 2017; Peñalba and Wolf 2020).

A fundamental requirement for trait evolution is underlying genetic variation (Falconer and MacKay 1996), and there is empirical evidence that recombination itself is evolving (reviewed in Ritz et al. 2017; Stapley et al. 2017). As early as the 1970s, experiments in *Drosophila melanogaster* have shown that recombination rates could be artificially selected (Chinnici 1971; Kidwell 1972). In addition, recombination rates and distribution have been observed to evolve as a result of strong selection on other traits, such as in experimental populations of

*Drosophila*, mice, and mustard, and in domesticated plants such as barley, tomato, and rye (Otto and Barton 2001; Dreissig et al. 2019; Fuentes et al. 2022; Schreiber et al. 2022). There is also evidence that recombination rates can evolve as a consequence of local adaptation, as observed in populations of *Drosophila pseudoobscura* (Samuk et al. 2020). Over longer evolutionary timescales, signatures of molecular evolution have been observed at meiotic genes in *Drosophila* and mammals (Brand et al. 2018; Dapper and Payseur 2019), and *PRDM9*, a locus associated with recombination hotspot positioning, is one of the fastest evolving genes in many vertebrates (Baker et al. 2017). There is also increasing evidence of genetic variation in recombination in more and more systems, indicating the *potential* for rates and distribution to evolve.

Understanding the genetic architecture of this variation - that is, which loci are involved, how many are involved, and their relative effects - is of key interest to the fields of molecular and evolutionary biology. It can not only identify molecular mechanisms underpinning recombination variation and make direct links to fertility and disease, but can also shed light on the evolutionary capacity and constraints on variation, and how this may in turn affect downstream evolutionary processes that are influenced by recombination. In this perspective, I explore what we have learned about the genetic basis of variation in recombination, and how this can help us to understand their evolution more broadly. I present an overview of molecular and evolutionary trade-offs and constraints associated with the rate and distribution of recombination. I then discuss progress made in quantifying recombination, and what we have learned about the genetic architecture of variation. I then finish by discussing the broader implications of these findings, and open questions for future research.

## Recombination: a trait characterised by molecular and evolutionary trade-offs.

The diversity of recombination presents an interesting conundrum, as the rate and distribution of recombination are characterised by both molecular and evolutionary trade-offs. Recombination can be beneficial from a molecular perspective, as the formation of crossovers is often critical to prevent aneuploidy and ensure the correct segregation of chromosomes into gametes, with most species having a minimum requirement of one obligate crossover per chromosome pair (Koehler et al. 1996; Hassold and Hunt 2001; Wang et al. 2015). This constraint means that chromosome number is often a major driver of the genome-wide recombination rate variation, where having more chromosomes leads to a higher minimum boundary of crossover rate (Pardo-Manuel de Villena and Sapienza 2001; Stapley et al. 2017; Brazier and Glémin 2022). However, there are also molecular costs; recombination can be associated with the formation of hundreds of DSBs in the genome, and their repair is directly mutagenic, with higher *de novo* mutation rates seen at DSB repair sites compared to the rest of the genome (Pratto et al. 2014; Halldorsson et al. 2019; Hinch et al. 2023; but see Liu et al. 2017). In addition to these benefits and costs, distribution of recombination events can be constrained by crossover interference (Muller 1916), a phenomenon where the formation of a crossover in one position reduces the probability of more crossovers forming nearby on the same chromosome (reviewed in Otto and Payseur 2019; von Diezmann and Rog 2021). Therefore, one may expect that these molecular trade-offs and constraints should limit the rate

> **Box 1: A brief history of understanding and measuring recombination.**
>
> The crossover process was first proposed more than a century ago, through experiments of Alfred Sturtevant on the co-inheritance of alleles controlling mutant phenotypes in *Drosophila melanogaster,* while he was a student in the laboratory of Thomas Hunt Morgan (Sturtevant 1913). Later, Harriet Creighton and Barbara McClintock validated this concept in cytological experiments in maize (Creighton and McClintock 1931). The process of gene conversion and the existence of hotspots was established through experiments in the fungi *Neurospora* and *Saccharomyces cerevisiae* in the 1960s (reviewed in Lichten and Goldman 1995; Lorenz and Mpaulo 2022), but the molecular mechanisms of meiotic recombination were not understood until the late 1980s, when the first genes involved in meiosis were identified in the *S. cerevisiae* (reviewed in Zickler and Kleckner 1999). There is now a rich understanding of the molecular mechanisms of recombination across eukaryotes, thanks to a wealth of work in model systems of the yeasts *Saccharomyces cerevisiae* and *Schizosaccharomyces pombe*, the thale cress *Arabidopsis thaliana*, the nematode worm *Caenorhabditis elegans*, the fruit fly *Drosophila melanogaster* and the house mouse *Mus musculus*, among others (reviewed in Arter and Keeney 2023).
>
> Our understanding of variation in recombination and its evolution has been tightly coupled with advances in technologies and methodologies to measure it (Table 1). Soon after recombination was discovered in *D. melanogaster* (Sturtevant 1913), it was shown to be plastic in response to environmental influences such as temperature (Plough 1917). Since the 1970s, variation in both the number and distribution of crossovers has been well characterised in *Drosophila* and mice (Henderson and Edwards 1968; Lyon 1976). From the late 1990s onwards, the emergence of pedigree-based studies with dense molecular marker data showed that there can be between-individual variation in crossover counts and crossover interference in humans and mice (Broman et al. 1998; Kong et al. 2004; Campbell et al. 2015). Another milestone was the development and application of sperm-typing methods to identify the positioning of recombination events, again supporting the idea that recombination can be punctuated in hotspots across the genome (Li et al. 1988; Jeffreys et al. 2001). This was rapidly followed by population-based studies investigating linkage patterns in whole-genome sequence data, which confirmed that recombination often occurs in hotspots of 1-10kb in many animals and plants (Myers et al. 2005; Hellsten et al. 2013; reviewed in Choi and Henderson 2015; Zelkowski et al. 2019).

of recombination to as few DSBs and crossovers as possible, yet widespread variation is still observed at the chromosomal level (Stapley et al. 2017; Fernandes et al. 2018; Brazier and Glémin 2022).

Another compelling explanation for the diversity of recombination rates is an evolutionary one, and there is a rich theoretical literature on this topic (reviewed in Otto and Lenormand 2002; Hartfield and Keightley 2012). Recombination has the benefit of providing a mechanism to purge deleterious mutations from genomes (Muller 1964; Kondrashov 1988) and can bring together beneficial variants at linked loci onto the same chromosome, increasing the speed and magnitude of responses to selection (Fisher 1930; Muller 1932; Felsenstein 1974). Similarly, recombination can mitigate the effects of selection at one locus interfering with selection at linked loci (i.e. Hill-Robertson interference; Hill and Robertson 1966). This is important in finite populations, where genetic drift can generate negative linkage disequilibrium between loci (Felsenstein 1974; Charlesworth et al. 2009; Roze 2021). However, there are also evolutionary costs to recombination: the same mechanisms can generate "recombination load", where beneficial variants at linked loci are uncoupled (Charlesworth and Charlesworth 1975; Charlesworth and Barton 1996) and, as described above, DSB repair is a major source of new and potentially deleterious mutations. Recombination can also induce genetic load through GC-biased gene conversion, as gene conversion events are more likely to be repaired with GC alleles rather than AT alleles, irrespective of whether they have a positive or negative effect on fitness (Galtier et al. 2009; Necşulea et al. 2011; Lachance and Tishkoff 2014).

The optimum rate and distribution of recombination will vary due to the relative importance of all of these factors, from the level of the chromosome to the level of individuals, sexes, and populations. It will not only depend on the strength of selection, genetic drift, and the deleterious mutation rate at a given time (Roze 2021), but also the direct impact of recombination on fitness traits, such as fertility and offspring viability (Hassold and Hunt 2001; Kong et al. 2004; Berg et al. 2010). The nature and complexity of these trade-offs may explain why variation in recombination is so pervasive, but our understanding has been limited by a lack of empirical investigation of selection on recombination itself. While theoretical studies have provided an important foundation for arguments on the evolution of recombination, they often make assumptions of the genetic architecture of recombination which may not be realistic. Therefore, better estimates of recombination rate and distribution, combined with a realistic understanding of its genetic architecture and impacts on fitness, are crucial to better understanding of how they are evolving.

## Overcoming the challenges of measuring recombination.

A first step in understanding variation in recombination is to quantify it, yet it is a highly challenging phenotype to measure (reviewed in Peñalba and Wolf 2020; Table 1). Recombination is a cellular process that takes place in different types of specialised reproductive tissues and cells (e.g., ovaries vs. testes/anthers, oocytes vs. spermatocytes), which can be difficult and invasive to sample directly, particularly if recombination is occurring during narrow developmental windows (e.g., in the foetal ovary in mammals). One way to overcome this challenge has been to infer recombination indirectly, either using genomic approaches to estimate recombination events present in gametes transmitted from parents to offspring (Kong et al. 2004; Dréau et al. 2019; Hinch et al. 2019), or using whole genome sequence data to infer recombination events over thousands to millions of generations (Auton and McVean 2007; Munch et al. 2014; Joseph et al. 2024). In truth, there is no "perfect" measure of recombination, with different approaches providing different insights but also different challenges (Table 1). Recent decades have seen major advances in measuring and understanding recombination (Box 1) and today, there are a variety of approaches to quantify it at different resolutions and timescales in a diverse range of species (see Table 1 and references therein). These advances have provided a much-needed foundation to understand the individual, environmental, and genomic factors associated with variation in recombination rates and distribution, often through a combination of multiple approaches. Furthermore, the ability to quantify recombination has also been crucial for advancing many evolutionary analyses, such as identifying and interpreting selective sweeps (Josephs and Wright 2016), inferring phylogenies and demographic histories (Li et al. 2019; Feng et al. 2023; Soni et al. 2024), and predicting population responses to selection (Battagin et al. 2016; Epstein et al. 2023).

## The genetic architecture of variation in genome-wide recombination rates.

As discussed above, a central goal in understanding the evolution and evolutionary potential of recombination variation has been to determine its genetic architecture. Specifically, we aim to determine the proportion of variation in recombination that is "heritable" (i.e., explained by additive genetic variation), identify the loci that contribute to heritable variation, and determine the strength of their effects. In addition, we aim to determine if these loci are either: (a) affecting genome-wide recombination rates (*trans*-acting); or (b) influencing recombination in their immediate vicinity (*cis*-acting) through inherited differences in chromatin accessibility, transposable elements, methylation, and/or structural variants such as inversions, which can suppress local rates of recombination (Stevison et al. 2011; Haenel et al. 2018; Zelkowski et al. 2019; Brazier and Glémin 2022).

In this section, I will focus on the genetic architecture of individual variation in genome-wide crossover rates, as to the best of my knowledge, that of individual variation in genome-wide gene-conversion rates has not been quantified. Similar principles will apply to understanding the genetic architecture of individual variation in recombination distribution, which I will address in the following section.

**GWAS reveals large effect loci for crossover rate in mammals.** A common approach to determine the genetic architecture of traits is to use genome-wide association studies (GWAS) to identify loci of moderate to large effects. Vertebrate populations with large pedigrees of individually genotyped individuals have been highly suitable for this, as they allow large numbers of individual recombination rates to be measured (See Table 1: Pedigree-based estimation) and for GWAS to be applied to the same individuals. Such studies have generally been limited to model and domestic vertebrates, such as humans, cattle, pigs, sheep, Atlantic salmon and chickens (Kong et al. 2014; Ma et al. 2015; Kadri et al. 2016; Petit et al. 2017; Weng et al. 2019; Johnsson et al. 2021; Brekke, Johnston, et al. 2022; Brekke, Berg, et al. 2022; Brekke et al. 2023), but are increasingly applied to long-term wild pedigrees, such as in Soay sheep, red deer, and house sparrows (Johnston et al. 2016; Johnston et al. 2018; McAuley et al. 2024). Genome-wide recombination rates in individual gametes (most often measured as crossover count) have been shown to be heritable in all of these systems. A striking observation is that all mammal studies to date show a consistent pattern of *trans*-acting loci with a large effect on the genome-wide recombination rate. Nearly all of these loci correspond to genes associated with meiotic processes, including double strand break initiation/repair and crossover designation (*RNF212*, *RNF212B*, *REC8*, *MEI1*, *MSH4*, and *PRDM9*[1], among others). Moreover, these loci often display sex-differences in their effects on recombination, where those that have a large effect in one sex (more often females) have little or no effect in the other sex (Kong et al. 2014; Johnston et al. 2016; Kadri et al. 2016; Johnston et al. 2018; Weng et al. 2019; Brekke, Johnston, et al. 2022; Brekke, Berg, et al. 2022; Brekke et al. 2023). The most extreme case has been observed in humans, where the locus *RNF212* can have sexually-antagonistic effects on recombination, where alleles conferring increased male recombination will decrease female recombination (and *vice versa;* (Kong et al. 2008; Kong et al. 2014)).

---

[1] I note here that the locus *PRDM9* is a special case, and I discuss its role in individual variation in more detail in the following section.

***Genetic variation in crossover rates in plants and insects.*** Recombination rate variation can also have a genetic basis in non-vertebrate systems. A GWAS in a large hybrid population of domestic and wild barley identified the locus *Rec8* as having a large *trans*-acting effect on crossover number and distribution (Dreissig et al. 2020). In *Arabidopsis thaliana*, quantitative trait locus mapping has identified naturally segregating loci associated with meiotic crossover rate (Ziolkowski et al. 2017; Lawrence et al. 2019). In insects, GWAS and family crosses in *Drosophila* species have identified loci with meiotic functions as being associated with *cis*- and *trans*-variation in crossover rates and distribution (Charlesworth et al. 1985; Cattani et al. 2012; Hunter et al. 2016), and a study of fine-scale recombination rates in the honey bee (*Apis mellifera*) also identified *trans*-acting, heritable variation in recombination rates (Kawakami et al. 2019).

***The genetic basis of crossover rate can be polygenic.*** Heritable variation in recombination rates is not always controlled by loci with large effects. In non-mammalian vertebrates, such as house sparrows, chickens, and Atlantic salmon, GWAS studies have shown that crossover rates can be polygenic (Weng et al. 2019; Brekke et al. 2023; McAuley et al. 2024). This means that they are controlled by many loci distributed across the genome that each have a relatively small effect on recombination rates. In such cases, it can be challenging to determine if polygenic variation is acting in *cis* and/or *trans*. *Cis*-acting variants will influence recombination in their immediate vicinity as described above, whereas *trans*-acting variants may have small effects on the cell environment, genome-wide chromatin structure, and/or general meiotic processes; both *cis*- and *trans*- effects will then cumulatively affect the genome-wide recombination rate. One approach has been to quantify the effect of loci on the recombination rate excluding the chromosome on which they reside, thereby removing any *cis* effects on recombination. In house sparrows, this approach indicated that whilst most polygenic variation appears to have a *trans* effect on crossover counts, the distribution of crossovers is likely driven by loci operating in both *cis* and *trans* (McAuley et al. 2024). Finally, polygenic effects can still differ between the sexes. In Atlantic salmon and house sparrows, the cross-sex additive genetic correlation is relatively low ($r_A \approx$ 0.11 and 0.30, respectively), meaning that females and males still have largely different genetic architectures of recombination rate, similar to what has been observed in mammals above (Brekke et al. 2023; McAuley et al. 2024).

## The genetic architecture of variation in recombination distribution.

The distribution of recombination events is primarily affected by the physical structure of chromosomes. During meiosis, DNA is tethered into chromatin loops along an axis structure, which is then integrated into the synaptonemal complex, a protein structure that holds homologous chromosomes in close proximity during recombination (reviewed in (Henderson and Bomblies 2021)). Larger loops and shorter axes/synaptonemal complexes are associated with lower crossover rates at the chromosome level, likely due to stronger effects of crossover interference over shorter physical distances (Lynn et al. 2002; Dumont and Payseur 2011; Ruiz-Herrera et al. 2017). The distribution of recombination can also vary relative to centromeres and telomeres, structural variants such as inversions, transposable elements, and other aspects of chromatin structure, such as accessibility and methylation (Stevison et al. 2011; Haenel et al. 2018; Zelkowski et al. 2019; Brazier and Glémin 2022). The distribution of recombination can be described in terms of variation at the broad-scale (i.e., at the chromosomal level), or at the fine-scale (i.e., in terms of its precise positioning). The causes and consequences of broad- and fine-scale variation are not necessarily mutually exclusive, and in both cases, there is evidence that genetic variants can contribute to observed variation.

**Broad-scale variation in crossover distribution.**

There is evidence that individual variation in the distribution of crossovers can be heritable, in terms of where crossovers tend to be positioned on chromosomes at the broad-scale, and the degree of crossover interference (i.e., how crossovers are positioned relative to one another). While such studies examining the genetic basis of these traits remain rare, they have been aided by improvements in the ability to measure individual crossover positioning and interference, which allow us to characterise a more quantitative picture of recombination distribution across the whole genome (Otto and Payseur 2019; Veller et al. 2019).

***There can be genetic variation in chromosome-level crossover positioning.*** One such quantitative measure is that of intra-chromosome genetic shuffling, $\bar{r}_{intra}$, which uses the positioning of crossover events to quantify the probability that two loci on the same chromosome become uncoupled during meiosis (Veller et al. 2019). Crossovers restricted to the chromosome ends will generate lower $\bar{r}_{intra}$ values, whereas central crossovers will generate higher $\bar{r}_{intra}$ values. Therefore, this metric can be used as a proxy for crossover positioning across the whole genome (e.g., if an individual has a lower $\bar{r}_{intra}$, then it is more frequently placing crossovers towards chromosome ends). It should be noted that $\bar{r}_{intra}$ may show some correlation with crossover rate, as (a) more crossovers can lead to more genetic shuffling (McAuley et al. 2024), and (b) the positioning of crossovers at the chromosome ends can allow more crossovers to occur on chromosomes when crossover interference is present (Dukić and Bomblies 2022). In addition, the use of this measure has provided insights into the realised effects of crossovers, particularly in the context of heterochiasmy. An example of this is in Atlantic salmon, where there are distinct differences in crossover distribution between females and males; males recombine almost exclusively in close proximity to the telomere, whereas females have higher crossover rates in proximity to the peri-centromere and will still recombine along the length of the chromosome (Brekke et al. 2023). As a result, females have around 1.6 times more crossovers than males, but have around 8 times higher genetic shuffling (Brekke et al. 2023). This means that the generation of novel linked diversity in this species due to recombination is mostly arising via females in this species. The measure $\bar{r}_{intra}$ has been shown to be modestly heritable and polygenic in Atlantic salmon and house sparrows (Brekke et al. 2023; McAuley et al. 2024). A recent analysis in domestic pigs has shown that variation in $\bar{r}_{intra}$ (independently of the crossover rate) can be associated with the large effect loci *MEI4*, *PRDM9* and *SYCP2*, which are associated with DSB formation, hotspot positioning, and the structure of the synaptonemal complex, respectively (Brekke et al. 2024).

***The distribution of recombination can have a genetic basis in domestic plants.*** Variation in crossover positioning has a genetic basis in domesticated rye, where a locus of major effect corresponding to the gene *ESA1* was associated with an increase in the size of low-recombining regions in domesticated lines, with no change in the genome-wide rate (Schreiber et al. 2022); the authors hypotheses that this has arisen though indirect selection to achieve more homogeneous populations for agricultural use. In tomatoes, fine-scale alterations in recombination in specific genomic regions has been observed between wild and domestic populations, with a loss of hotspots associated with selective sweeps (Fuentes et al. 2022), and in barley, there is evidence that domestication has led to reduced recombination rates in interstitial chromosomic regions, but higher rates in distal regions (Dreissig et al. 2019).

***There can be genetic variation in crossover interference.*** The strength of crossover interference can also vary across chromosomes, and in turn impact recombination rates and distribution (reviewed in Otto and Payseur 2019). Individual variation in crossover interference has been shown to be heritable in cattle and pigs, where it was associated with variants at the loci *NEK9* and *RNF212*, respectively (Wang et al. 2016; Brekke et al. 2024). Similarly, an investigation of individual variation in synaptonemal complex lengths in mice identified *RNF212* as a candidate locus (Wang et al. 2019); this may function to increase crossover interference by shortening the physical distance in which more crossovers can be placed. The identification of *RNF212* affecting crossover interference and potentially the synaptonemal complex length in mammals is interesting, as there is emerging evidence that the dosage and meiotic behaviour of *Hei10* (a locus in the same conserved family of E3 ubiquitin ligases, and with similar function to *RNF212* during meiosis) is highly associated with crossover interference in *Arabidopsis thaliana* (Morgan et al. 2021; Girard et al. 2023). However, studies on the genetic architecture of crossover interference remain challenging, as it relies on many observations of two or more crossovers on the same chromosome (which are less frequent at the genome-wide scale). This means that the sample sizes required to accurately characterise crossover interference remain difficult to achieve, with the exception of species where large numbers of gametes can be investigated cytologically, or in large pedigrees with many offspring per individual as observed in domesticated systems.

**Fine-scale variation in recombination distribution.**

The distribution of recombination events can differ at the fine-scale along chromosomes. In many species, DSB and crossover events often occur in recombination hotspots of between 1-10kb in width (Myers et al. 2005; Choi and Henderson 2015), although there are exceptions (see below). Fine-scale variation in recombination is challenging to measure at the individual level; therefore much of the work in this section has relied on population- and phylogeny-based estimation, as well as findings from ChIP-seq and gamete sequencing approaches (Table 1).

***Ancestral hotspots are evolutionarily stable and enriched at functional elements.*** In most species with hotspots (and most likely the ancestral state), hotspots tend to occur around functional elements such as gene promoter regions, and are often associated with regions of open chromatin, nucleosome depletion, and reduced DNA methylation (Lichten and Goldman 1995; Brachet et al. 2012; Zelkowski et al. 2019; Lian et al. 2022). This is likely to be because these factors improve the ability of meiotic proteins, such as SPO11, to access and bind to DNA in order to initiate DSB formation (Pan et al. 2011). The positions of these ancestral hotspots can be conserved over long evolutionary time periods (i.e., millions of years), as demonstrated in mammals (Joseph et al. 2024), birds (Singhal et al. 2015), and yeast (Tsai et al. 2010). The distribution of hotspots can also be correlated with genomic features associated with broad-scale variation above, including transposable element content, structural variation, and histone H3 lysine K4 trimethylation marks (H3K4me3; (Kent et al. 2017; Morgan et al. 2017; Zelkowski et al. 2019)). Fine-scale recombination rates are also often positively correlated with GC-content (e.g. as in mammals), most likely as a consequence of GC-biased gene conversion (Duret and Galtier 2009).

***PRDM9-mediated hotspots evolve rapidly.*** One of the most notable discoveries in the field of recombination is that of *PRDM9*, a rapidly-evolving locus that that determines recombination hotspot positioning in many vertebrate species (Baudat et al. 2010; Myers et al. 2010; Parvanov et al. 2010; Baker et al. 2017). Unlike the ancestral hotspots above, PRDM9-mediated hotspots tend to be directed away from functional elements, and show remarkably little conservation between closely-related species (Berg et al. 2010; Baker et al. 2015; Stevison et al. 2016; Wooldridge and Dumont 2023). One key feature of PRDM9 is its zinc-finger (ZF) array, which binds to specific sequence motifs throughout the genome (Ségurel et al. 2011). Mutations in the ZF array can change the recognised sequence motifs to which it binds, essentially leading to the immediate loss and gain of recombination hotspots (Davies et al. 2016). Another feature of this system is that if there is asymmetry in hotpot sequence motifs (i.e. if a hotspot is heterozygous), then DSBs will form at the allele more likely to be bound by PRDM9, and repaired with the allele less likely to be bound by PRDM9 (Myers et al. 2008). This leads to a rapid loss of hotspots from the genome, referred to as the "hotspot paradox" (Boulton et al. 1997; Coop and Myers 2007). This may result in an "arms race" scenario to replenish hotspots through selection for new sequence motifs (Ubeda and Wilkins 2011; Latrille et al. 2017), and indeed, in species where *PRDM9* is likely to be functional, it is one of the most rapidly evolving genes in the genome (Baker et al. 2017). The diversity of the *PRDM9* ZF array within species can be remarkable and may reflect variation in their binding affinities to different motifs; over 150 alleles have been identified in wild mice (Vara et al. 2019; Wooldridge and Dumont 2023), 69 alleles in humans (Alleva et al. 2021), and even 22 alleles identified in 19 corn snakes in a single study (Hoge et al. 2024).

There is evidence that *PRDM9* alleles may also affect the genome-wide rate of recombination. As mentioned in the previous section, variants at *PRDM9* have been associated with genome-wide crossover rates in humans, cattle, and pigs (Kong et al. 2014; Ma et al. 2015; Brekke, Berg, et al. 2022). In these cases, *PRDM9* alleles may affect the genome-wide rate if they have differences in their binding efficiency, and/or if different alleles are binding to more common or more rare sequence motifs. *PRDM9* has also been identified as a "speciation gene" in mice, where asymmetry in DSB distribution at heterozygous binding sites may lead to sterility in hybrid males (Mukaj et al. 2020; Davies et al. 2021); indeed, *PRDM9* is the only speciation gene identified in mammals to date. Finally, *PRDM9* has also been implicated as a risk factor for disease-associated genome rearrangements in humans (Berg et al. 2010), providing another avenue by which *PRDM9* may be under selection.

***Ancestral and PRDM9-mediated hotspots can co-exist.*** At the time of writing, insights into the ubiquity and relative importance of PRDM9-mediated hotspots are still emerging. PRDM9 was first shown to be the major driver of hotspot positioning in humans and mice (Baudat et al. 2010), and there is now evidence that it is associated with hotspots in nearly all mammals, some teleost fish, turtles, snakes, and lizards (Baker et al. 2017; Schield et al. 2020; Hoge et al. 2024; Raynaud et al. 2024); there is also emerging evidence that *PRDM9* may direct hotspot positioning in some insects (Everitt et al. 2024). However, *PRDM9* function has been lost in some groups, such as canids, birds, crocodiles and amphibians (Baker et al. 2017), which have reverted back to the stable, ancestral hotspots enriched at functional elements (Singhal et al. 2015). Indeed, reversion to ancestral hotspots has been experimentally confirmed as a direct consequence of knocking out *Prdm9* in mice (Brick et al. 2012). Until recently, it was generally considered that *PRDM9* was the overwhelming driver of recombination hotspot positioning in species where it is functional, based on findings from humans and mice. However, more studies are beginning to show that the high fidelity of recombination to PRDM9-mediated hotspots observed in these species may be the exception. A recent investigation of hotspots in 52 mammal species showed that many species in fact use both PRDM9-mediated and PRDM9-independent (i.e. ancestral) hotspots (Joseph et al. 2024), and that this is also likely to be the case in rattlesnakes and corn snakes (Schield et al. 2020; Hoge et al. 2024).

***Some species lack hotspots.*** It should be noted that some species do not have recombination hotspots, such as *Drosophila spp.* (Chan et al. 2012; Smukowski Heil et al. 2015) and the nematode *Caenorhabditis elegans* (Kaur and Rockman 2014). Nevertheless, transgenic loss-of-function experiments in these species have identified meiosis genes that can modify crossover distribution. These include the *mei-217/-218* locus in *Drosophila melanogaster* and *D. mauritiana,* which alters rates within defined genomic intervals (Brand et al. 2018), and the *rec-1* locus in *Caenorhabditis elegans,* where the loss-of-function mutant leads to more homogenous distribution of recombination across chromosomes (Parée et al. 2024). While the mechanisms and evolutionary implications of this absence of hotspots remain unknown (Zelkowski et al. 2019), it has been suggested by Baker et al. (2023) that a common feature of species without hotspots is that they pair homologous chromosomes independently of DSB formation, unlike species with hotspots, where homologue pairing relies on DSB and recombination formation (reviewed in Keeney et al. 2014).

## The genetic basis of recombination: open questions and future directions.

This perspective has provided a snapshot of developments in understanding the genetic architecture of recombination variation over recent decades. One clear message is understanding the vast variation in recombination is a complex and dynamic challenge. This is because recombination is a phenotype of the genome, affected by not only the structure of the genome, but also the genetic variation encoded within it, both of which can evolve rapidly. This is compounded by the fact that recombination does not just vary across the genome, but also between individuals, sexes, populations and species, and can be plastic relative to individual and environmental effects. Finally, recombination variation can have direct impacts on fertility and health, but also indirect impacts on the ability for populations to adapt in response to selection, and potentially, could affect the very survival of those populations.

Ultimately, there is no one-size-fits-all canonical model of recombination evolution. Current and future progress will continue to rely on developing technologies and the synergistic and inter-disciplinary research investigating different facets of this problem within diverse biological systems. To end this perspective, I consider the immediate and future challenges of the field by presenting open questions on the genetic architecture and evolution of variation in recombination. This list is neither exhaustive nor mutually exclusive, and will likely reflect my own biases and knowledge; therefore, I urge the reader to explore the reviews cited throughout this perspective to gain a full picture of the challenges that remain.

**What is the relationship between fine-scale genome dynamics and recombination variation?** Improvements and falling costs of long-read DNA sequencing are providing high quality genome assemblies and annotations (including structural variation) which are improving the accuracy of sequence-based analyses outlined in Table 1. Similarly, these advances may permit better characterisation of individual variation in DSB formation and/or gene-conversion events, creating the opportunity to determine if variation has a similar or different genetic architecture to e.g. crossover count, positioning, and interference. A promising technological advance is the ability to investigate genome dynamics at the single cell level during meiosis, including chromatin accessibility (e.g. ATAC-Seq), gene expression (e.g. RNA-Seq), and potentially chromosome conformation capture (e.g., Hi-C), which may provide more insight into the specific mechanisms of variation in recombination (reviewed in Peng and Qiao 2021; Jovic et al. 2022). For example, these approaches may elucidate relationships between genome-wide gene expression and recombination distribution, and if particular gene pathways could be more prone to harbour recombination hotspots if they are transcribed and expressed in meiotic cells.

**What are the molecular mechanisms by which genetic variation affects recombination?** A huge body of research on PRDM9 has led to a detailed understanding of the mechanisms by which genetic variation at this locus impacts recombination distribution (Grey et al. 2018), but this level of insight is the exception rather than the norm. Experimental work in model systems has established the molecular functions of meiotic loci associated with genome-wide variation in crossover count, positioning and interference, such as *RNF212*, *Hei10*, and *MEI4* (Arter and Keeney 2023). However, the precise mechanisms by which segregating alleles at these loci affect variation in recombination (e.g., through differences in protein coding regions, gene expression, and/or interactions with the genome) are less understood. New insights may be gained from using a combination of approaches, such as: (a) fine-mapping and functional prediction of candidate causal mutations in sequence data; (b) targeted quantification of single-cell gene expression in gametocytes from individuals with different recombination alleles; and/or (c) functional validation through gene editing, where feasible.

**Are there direct associations between genetic variation of recombination and fitness?**
There is some evidence of long-term selection on recombination (e.g. (Dapper and Payseur 2019; Samuk et al. 2020), but there is less evidence of direct associations between genetic variation in recombination and individual fitness (i.e. fertility and viability). There are some rare examples at the phenotypic level: in a human population, individuals with higher oocyte crossover rates tended to have more offspring (Kong et al. 2004); but conversely, an experiment imposing selection for high fertility in mice led to a correlated reduction in recombination (Gorlov et al. 1992). A major hurdle for this question is that direct measurement of recombination can be invasive, and indirect pedigree-based measures cannot estimate the recombination rates of individuals that did not have offspring (Table 1). A solution may be to leverage quantitative genetic approaches, including: estimating genetic correlations between recombination rates and fitness (Kruuk 2004); genomic prediction of recombination rates in individuals without offspring (e.g. Hunter et al. 2022)); and/or investigating direct associations between major effect loci and fitness traits (e.g. Johnston et al. 2013)). It should be noted that selection in recombination may be transient and rarely observed, or that variation in recombination rate can exist with no apparent fitness cost (e.g. as observed in *A. thaliana*; Fernandes et al. 2018).

**Why does genetic variation for recombination persist despite strong selection?** One factor that remains unexplained is why some large-effect loci consistently maintain genetic variation within populations. An notable example of this is *RNF212*, which is essential for the crossover designation process (Reynolds et al. 2013), yet consistently has a large effect on crossover rate variation in nearly every mammal study discussed above. This is particularly puzzling given that many of these systems are domestic mammals under strong selection, which is hypothesised to favour increased rates of recombination in order to overcome Hill-Robertson interference (Otto and Barton 2001). Therefore, we may predict that high recombination alleles should become fixed in these populations. However, despite initial assertions that they had (Burt and Bell 1987), recombination rates have more likely not increased with domestication in mammals (Muñoz-Fuentes et al. 2015). Indeed, given that recombination contributes relatively little to the shuffling of mammal genomes relative to independent assortment of chromosomes (Veller et al. 2019), recombination has a relatively weak impact on genetic improvement as chromosome numbers increase (Battagin et al. 2016; Gonen et al. 2017). In contrast, some plant species have shown evidence of evolution on recombination under domestication, including rye, tomato and barley as described above (Dreissig et al. 2019; Fuentes et al. 2022; Schreiber et al. 2022). This may be a consequence of crop plants having large non-recombining regions, meaning that modifying the distribution of recombination will be particularly beneficial for generating novel linked genetic variation (Epstein et al. 2023).

**Why does recombination differ between female and male gametes?** Sex-differences in recombination are common across eukaryotes and can vary in degree and magnitude even between closely related species, but there is little or no support for different evolutionary theories of why this is the case (reviewed in Lenormand and Dutheil 2005; Sardell and Kirkpatrick 2020). However, there is increasing evidence that female and male recombination variation can have different genetic architectures in vertebrates, which may allow some degree of evolutionary independence (Kong et al. 2014; Brekke et al. 2023; McAuley et al. 2024).

Furthermore, evolutionary theories rarely consider the molecular differences between female and male meiosis, including differences in developmental timing, synaptonemal complex length, gene expression and chromatin structure. Future work addressing this question will benefit from investigation of differences in fine-scale genome structure, including hotspot positioning, to determine the relative importance of evolutionary and molecular drivers.

**How does the genetic architecture of recombination affect its evolution?** As shown above, the genetic architecture of recombination can be relatively simple and driven by a small number of large effect loci, or it can be polygenic and driven by many loci of small and varying effects throughout the genome (or, somewhere in between). This genetic architecture can affect the speed and degree at which recombination can evolve. Simple architectures may permit rapid evolution, but with less scope to fine-tune the optimal level of recombination; whereas polygenic architectures may maintain genetic variation due to having a large mutational target for the introduction of new variants affecting recombination (Rowe and Houle 1996), the distribution of selection coefficients over many loci (Sella and Barton 2019), or pleiotropy and linkage with other traits under differing selection (Teplitsky et al. 2009; Ruzicka et al. 2019). One approach to addressing this question is to adapt theoretical studies to model more realistic genetic architectures identified above, in order to make better predictions on the evolutionary drivers and consequences of recombination variation.

## Acknowledgements

I sincerely thank two anonymous reviewers for their detailed and helpful comments. I thank Deborah Charlesworth, Julien Joseph, Raphaël Mercier, Steven Dreissig, Peter Keightley, Simon Martin, Bertrand Servin, John McAuley, and Cathrine Brekke for feedback and discussions on all aspects of this topic. I thank Aimeric Blaud for support throughout the writing process.

## Funding

I am funded by a Royal Society University Research Fellowship (URF/R/211008).

## Data Availability

There is no data associated with this manuscript.

## Referenced Literature

**Table 1. Summary of current methods to quantify variation in recombination rate and distributions.** See Peñalba and Wolf (2020) for a detailed review on different approaches. CO and GC refer to crossovers and gene conversion events, respectively. References to methods and/or empirical examples are provided below the table. Sample size indicates the minimum number of individuals required for meaningful characterisation of variation.

| Method | Description | Pros | Cons | Sample Size |
|---|---|---|---|---|
| Cytogenetic estimation: | Directly visualises chromosomes in gametocytes using immunostaining of meiotic proteins to identify COs, often targeting foci of DNA mismatch repair protein MLH1. Can identify number and distribution of DSBs and the length of the meiotic axis/synaptonemal complex (e.g. targeting RAD51, SYCP3). | Direct observation of DSB and CO rates. Physical positions are determined e.g. in µm. | Requires invasive sampling of gametocytes & often limited to males. Cannot give fine-scale positions relative to sequence features. | ≥1-10s individuals |
| Pedigree-based estimation: | Integrates pedigree and genetic marker information (e.g. SNPs) to identify marker pairs separated by recombination in gametes transmitted from parents to offspring. Can estimate recombination in (a) *individuals*, using information on recombination positions in gametes; and (b) *populations*, creating linkage maps measured in centiMorgans (cM), where 1cM is a 1% chance that two loci are separated by a CO event per meiosis. | Uses existing data from genotyped pedigrees. Quantifies individual variation. Can potentially identify GC-events in whole genome data. | Requires large sample sizes to capture enough COs. Resolution of recombination positions limited by marker density. | ≥100-1,000s individuals. |
| Gamete sequencing: | Sequences single- or pooled-gamete samples to identify recombination positions based on the deviation from parental or consensus allele frequencies within the same gamete and/or on the same sequencing read. | High precision of potential recombination positions within single individuals. | Often limited to male gametes due to ease of sampling. High sequencing costs. | ≥1-10s individuals |
| Chromatin immunoprecipitation sequencing (ChIP-Seq): | Isolates gametocytes and sequences genomic locations where specific proteins are bound to DNA using immunoprecipitation. DSB positions can be mapped by targeting proteins that initiate meiotic DSB formation (e.g., RAD51, DMC1). | Identifies the specific sites of DSB formation within single samples. | Requires invasive sampling of gametocytes & often limited to males. Difficult to verify if DSBs are repaired by CO or GC. | ≥1-10s individuals |
| Population-based estimation: | Uses whole genome sequence data to estimate the population-scaled recombination rate (ρ), based on patterns of linkage disequilibrium and the coalescent model. | Estimates fine-scale, sex-averaged recombination patterns over 100-1000s of generations. | Affected by demography and selection, but new methods (e.g. neural networks) may overcome this. Cannot distinguish CO and GC-events. | ≥10-30 individuals, including outgroups. |

| Phylogeny-based estimation: | Two main approaches: 1) leverages incomplete lineage sorting (ILS) in phylogenies to infer recombination breakpoints in ancestral branch of two extant sister species; or 2) uses the footprint of GC-biased gene conversion in substitution patterns to quantify relative fine-scale recombination rates on terminal branches of phylogenetic trees. | Fine-scale estimates averaged over long periods of time (>millions of years) using few genomes. | 1) Requires substantial ILS between targeted species. 2) Requires substantial GC-biased gene conversion in the tree. | Genomes from >4 or 3 closely related species, respectively. |
|---|---|---|---|---|

References : Cytogenetics (Malinovskaya et al. 2018; Peterson et al. 2019). Pedigree-based estimation (Kong et al. 2004; Brekke, Berg, et al. 2022; McAuley et al. 2024). Gamete sequencing (Dréau et al. 2019; Hinch et al. 2019; Xie et al. 2023). ChIP-Seq Immunoprecipitation (Smagulova et al. 2011; Tock et al. 2021; Lian et al. 2022). Population-based estimation (Auton and McVean 2007; Dapper and Payseur 2018; Adrion et al. 2020; Bascón-Cardozo et al. 2024). Phylogeny-based estimation (Munch et al. 2014; Joseph et al. 2024).